\documentclass{epl_submit}
\usepackage{graphicx,cite}

\def\eqref#1{(\ref{#1})}

\newcommand{\power}[3]{\left({{#1} \over {#2}}\right)^{#3}}

\newcommand{\twid}{\sim}

\newcommand{\casesshortii}[4]
{\left\{
\begin{array}{ll}
#1\ \ ,\gap & #2 \\  #3\ \ ,\gap & #4 
\end{array}%
\right.
}

\newcommand{\intinf}{\int_0^{\infty} }

\newcommand{\gap}{\hspace{.4in}}

\newcommand{\gt}{\rightarrow}

\newcommand{\period}{\ \ .}
\newcommand{\comma}{\ ,\ }

\newcommand{\lsim}{\,\stackrel{<}{\scriptstyle \sim}\,}

\newcommand{\ignore}[1]{}

{\end{list}}

{\end{list}}

{\ \\ XXXXX \hfill XXXXX #1 XXXXX\begin{equation}\label{#1}}%
{\end{equation}}

{\ \\ XXXXX \hfill XXXXX #1 XXXXX\begin{eqnarray}\label{#1}}%
{\end{eqnarray}}

\newenvironment{bequation}[1]%
{\begin{equation}\label{#1}}%
{\end{equation}}

\newenvironment{beqnarray}[1]%
{\begin{eqnarray}\label{#1}}%
{\end{eqnarray}}

\newcommand{\ie}{i.\,e.~}

{\begin{center} \section*{#1} \end{center}}

{\end{figure}} 

{\end{figure}}

{\end{figure}}

\newenvironment{eq}[1]%
{\begin{bequation}{#1}}{\end{bequation}}

{\begin{beqnarray}{#1}}{\end{beqnarray}}

\newcommand{\taus}{\tau_s}
\newcommand{\tsat}{t_{\rm sat}}

\newcommand{\smin}{s_{\rm min}}

\newcommand{\Omegat}{\Omega_t}

\newcommand{\tauN}{\tau_N}

\newcommand{\Gammabound}{\Gamma_{\rm bound}}

\newcommand{\Gammaboundfinal}{\Gamma_{\rm bound}^{\rm final}}

\newcommand{\Rtotal}{{\cal R}_{\rm total}}

\newcommand{\omegat}{\omega_t}
\newcommand{\omegatdot}{\dot{\omega}_t}

\newcommand{\Omegatdot}{\dot{\Omega}_t}

\newcommand{\Dsmaller}{D^{<}}
\newcommand{\Dgreatercrit}{D^{>}_{\rm crit}}
\newcommand{\Dgreater}{D^{>}}

\newcommand{\tfinal}{t_{\rm final}}

\newcommand{\sstar}{s^*}

\newcommand{\zstar}{z^*}

\newcommand{\Gammatot}{\Gamma_{\rm tot}}
\newcommand{\Gammainner}{\Gamma_{\rm inner}}
\newcommand{\cinner}{c_{\rm inner}}

\newcommand{\fcrit}{f_{\rm crit}}

\newcommand{\Ne}{N_{\rm e}}

\newcommand{\Nemelts}{N_{\rm e}^{\rm melts}}

\newcommand{\phinet}{\phi_{\rm net}}

\title{The Slowly Formed Guiselin Brush}
\author{Ben O'Shaughnessy  \and Dimitrios Vavylonis}
\institute{Department of Chemical Engineering, Columbia University,
New York, NY 10027, USA}


\begin{document}

\maketitle

\begin{abstract}
We study polymer layers formed by irreversible adsorption from a
polymer melt.  Our theory describes an experiment which is a ``slow''
version of that proposed by Guiselin [{\em Europhys. Lett.}, {\bf 17}
(1992) 225] who considered instantaneously irreversibly adsorbing chains
and predicted a universal density profile of the layer after swelling
with solvent to produce the ``Guiselin brush.''  Here we ask what
happens when  adsorption is {\em not} instantaneous.  The classic
example is chemisorption.  In this case the brush is formed slowly and
the final structure depends on the experiment's duration,
$\tfinal$. We find the swollen layer consists of an inner region of
thickness $z^* \sim \tfinal^{-5/3}$ with approximately constant
density and an outer region extending up to height $h \twid N^{5/6}$
which has the same density decay $\twid z^{-2/5}$ as for the Guiselin
case.
\end{abstract}


\section{Introduction}

\begin{figure}[t] 
\onefigure[width=\textwidth]{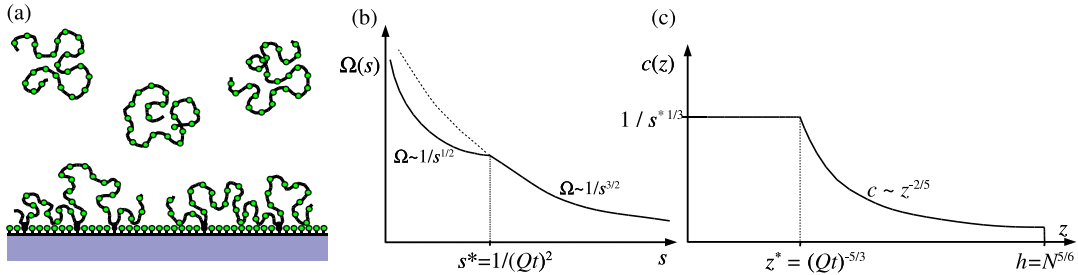}
\caption{
(a) Schematic of polymer melt containing 
reactive chains in
contact with a functionalised surface.  As reactions proceed,
irreversible bonds form (black groups) and a layer consisting of loops
and tails develops.  Experimentally the degree of polymer
functionalization, $f$, and fraction of reactive chains, $X$, can be
varied (here we mainly consider $f=X=1$).  Bond formation
is slow and requires a very large number of monomer-surface
collisions.  
(b) Distribution of chemisorbed loop sizes after time $t$.  Loops
smaller than $\sstar(t) \approx 1/(Qt)^2$ descend unhindered by others
and follow the free chain power law $\Omegat \twid s^{-1/2}$. The
distribution of big loops ($s>\sstar(t)$) is frozen in time, $\Omegat
\twid s^{-3/2}$. Total grafting density $\approx 1/{\sstar}^{1/2}$. As
time proceeds the dividing line shifts to the left and the future
distribution is shown (dotted line). For a given loop size
$s<\sstar(t)$, the density will continue to grow until the critical
value is reached and then halt. If uninterrupted the final loop
distribution, a single $s^{-3/2}$ power, is reached after $t\approx
Q^{-1}$. (c)  Density profile of the swollen chemisorbed layer.  The
profile exhibits two regions, separated by $z=\zstar$. The inner
region has approximately constant density while the density in the
outer region decays as $z^{-2/5}$.  The longer the
melt chemisorption is allowed to persist, the thinner is the inner
region.  For experimental times exceeding $Q^{-1}$ it
shrinks to zero.
}
                                                \label{scheme}
                                                \end{figure}

Several years ago Guiselin \cite{guiselin:irrev_ads} proposed an
experiment to study irreversible polymer adsorption. In its simplest
form, a melt is exposed to a surface so attractive to the polymer
chains that they adsorb instantaneously and irreversibly. This freezes
in melt chain configurations, including the size distribution of
surface loops (see fig. \ref{scheme}(a)). The adsorbed layer is then
swollen with solvent, washing away unattached chains.  Guiselin
predicted that the resulting interfacial structure, which has come to
be known as the ``Guiselin brush,'' has a universal density profile,
$c(z) \twid z^{-2/5}$ in good solvent.  Neutron scattering
\cite{auvray:irrev_ads_neutron,auvray:irrev_ads_concentrated_solution}
and neutron reflectivity
\cite{ben:pmma_adsorption,marzolin:eng_graft_and_guiseling_brush}
studies have indicated density profiles both consistent
\cite{auvray:irrev_ads_concentrated_solution,marzolin:eng_graft_and_guiseling_brush}
and inconsistent \cite{ben:pmma_adsorption} with this prediction.

In this letter we study the same irreversible melt adsorption
processes as did Guiselin, but we ask: what replaces the Guiselin
brush if the adsorption is {\em not} instantaneous?  The most
important example of this is {\em chemisorption} where functionalized
chains develop polymer-surface bonds which are usually
irreversible. This arises in various technologies where polymers are
attached to solid surfaces to permanently modify surface properties
\cite{kraus:book_reinforcement_elastomers,wu:polymer_iface_adhesion_book,%
edwards:review_filler_reinforcement} (fig. \ref{scheme}(a)).  For
example in fiber-reinforced thermoplastics, strong polymer-fiber
interfaces are frequently created by chemisorption of polymers onto
the fibers after functionalization of fiber surfaces with coupling
agents \cite{edwards:review_filler_reinforcement,cleggcollyer:book,%
pireaux:polymer_solid_ifaces_book}.  Another related class involves
reinforcement of immiscible polymer interfaces by chemical reactions
between multi-functionalized chains at the interfaces
\cite{creton:kramer:iface_fracture_review}.

Unlike the physisorption processes studied by Guiselin, chemisorption
is extremely slow (microscopically speaking) in that 2 mutually
reactive groups must collide millions of times, typically, before
bonding \cite{thon:rateconstants_book}. Hence surface loops usually have time to
explore all configurations before further reactions constrain their
motion.  As we will see, this leads to a different surface loop
structure and a modified version of the Guiselin brush after solvent
swelling.

Due to screening, polymer melt statistics are ideal. Now the
probability a random walk originating from a surface never re-contacts
that surface after $s$ steps is $\twid 1/s^{1/2}$. Its derivative
gives the surface loop distribution $\omega(s) \twid 1/s^{3/2}$ in a
melt. In Guiselin's experiment this is instantaneously frozen in.  To
see why slow chemisorption produces different $\omega(s)$ consider
first an unentangled melt of chains each comprising $N$ chemically
reactive units.  Chains within a coil radius $N^{1/2}$ of the surface
make many surface contacts (taking monomer size as unity). Now
reactions are switched on and the surface density (per site of size
unity) of bonded monomers, $\Gammabound(t)$, starts increasing from
zero.  After time $\tauN$, of order 1 polymer-surface bonds per chain
in this slab will have been created, \ie
                                                \begin{eq}{gammabound-taun}
\Gammabound (\tauN) \approx 1/N^{1/2} \comma\gap
                        \tauN = 1/(Q N^{1/2}) \period
                                                                \end{eq}
Here $Q$ is the local reaction rate {\em given} a polymer group
contacts a surface site (all of which are assumed reactive for
simplicity). Eq. \eqref{gammabound-taun} equates the reacted fraction
of surface sites, $Q\tauN\ll 1$, to the number of chains per site in
the slab, $1/N^{1/2}$. 


\section{Single Chain Adsorption: early stages} 

At this stage the slab contains surface-grafted loops (and tails) of
length $\approx N$.  Let us follow how one of these mother loops
gradually adsorbs down onto the surface (tails behave similarly) from
the moment of its creation. To begin, we ignore interference from
other chains.  Because $Q$ is small, the reaction rate of the loop's
$s^{\rm th}$ unit is proportional to its equilibrium surface contact
probability. For small $s$, this is independent of $N$,
                                                \begin{eq}{duck}
k(s|N) \gt k(s) \approx Q/s^{1/2} \comma\gap (s\ll N)  \period
                                                                \end{eq}
Hence the loop lifetime, the inverse of its total reaction rate
$\Rtotal \approx \int_1^N ds k(s|N)$, is identified with $\tauN$ of
eq. \eqref{gammabound-taun}. The loop evolution kinetics are
\cite{ben:chemiphysi_euro,ben:chemiguiselin,shafferchakraborty:pmma_chemisorption_kinetics}
                                                \begin{eq}{loop-kinetics}
\omegatdot(s) = 2 \int_s^N ds' k(s|s') \omegat(s') 
                                - \omegat(s) \int_0^s ds' k(s'|s) \period
                                                                \end{eq}
In this section $t$ denotes time after the mother loop was created,
and $\omegat$ describes an ensemble of mother loops plus daughters.
For small times, $t\ll\tauN$, to leading order there is just the
single mother loop, $\omegat(s)\approx \delta(s-N)$.  Substitution
into the first term in eq. \eqref{loop-kinetics}, describing creation
of $s$-loops, immediately gives
                                                \begin{eq}{curried-rabbit}
\omegat(s) = Qt/s^{1/2}\comma\gap \Dsmaller(s) = t/\taus
        \comma\gap \taus=1/(Q s^{1/2})\gap (s\ll N) \comma
                                                                \end{eq}
where $\taus$ is the lifetime of an $s$-loop. Note that the number of
loops shorter than $s$, $\Dsmaller(s)\equiv \int_0^s \omegat$, is very small
since $t\ll\tauN\ll\taus$ for these small loops. We conclude that in
addition to the single mother $N$-loop there are of order $t/\tauN$
smaller loops following a $1/s^{1/2}$ power law, distinct from the
$3/2$ decay in the fast Guiselin brush.  Note our argument neglected
the second (loop loss) term: substituting the power law into
eq. \eqref{loop-kinetics}, one finds it is self-consistently much
smaller (of relative order $t/\taus$) than the source term.


\section{Single Chain Adsorption: Collapse} 

For times greater than its lifetime $\tauN$, the mother loop will have
come down, spawning 2 daughter loops each of size $\approx N/2$ since
the total reaction rate $\Rtotal$ is dominated by $s$ of order $N$.
The daughters will in turn spawn 2 roughly equal granddaughters, and
so on. As this process iterates itself, more and more loops are
generated; the characteristic size after time $t$ is $\sstar =
1/(Qt)^2$.  Bigger loops, $s>\sstar$ have already come down ($\taus<t$)
whereas smaller loops have yet to be created ($\taus>t$). We can think
of the mother as having multiplied into $L(t)= N/\sstar\twid t^2$
offspring of equal size $\sstar$.

For much smaller loops, the source term in the kinetics
eq. \eqref{loop-kinetics} is now dominated by $s'\approx \sstar$ and
gives $\omegatdot(s) \approx k(s) \int_s^\infty ds'\omegat(s') \approx
k(s) L(t)$. Thus the loop distribution is
                                                \begin{eq}{loop-distribution}
\omegat(s\ll\sstar) \approx {L(t) \over \sstar} \power{\sstar}{s}{1/2}
                                \comma\gap 
\omegat(s\gg \sstar) \gt 0 \period
                                                                \end{eq}

We refer the reader to ref. \cite{ben:chemiguiselin} for detailed
analysis of the kinetics, eq. \eqref{loop-kinetics}, which are able to
justify the simple arguments presented here.  These kinetics do indeed
produce a distribution with the above features, \ie sharply cut off
above $\sstar$ and with $1/s^{1/2}$ behaviour for small $s$.  The
crucial point is that single loop adsorption is a {\em homogeneous
collapse} in which all parts of the mother loop come down essentially
at the same time.  Generally, the class of adsorption kinetics is
governed by the contact exponent $\theta$ where $k(s)\twid
s^{-\theta}$. Three classes are identified in
refs. \cite{ben:chemiphysi_letter,ben:chemiphysi_euro}:
zipping, for $\theta>2$; accelerated zipping for $1<\theta<2$; and
homogeneous collapse, for $\theta<1$. The present case, $\theta=1/2$,
is collapse; because $k(s)$ decays slowly, adsorption kinetics are
dominated by distant units of order the current loop size.


\section{Many Chains} 

Our single chain collapse description neglects interference from other
chains, appropriate when only a small fraction of chains are reactive.
However, when all chains simultaneously attempt to collapse as in the
present situation, grafted loop surface densities rapidly reach
criticality. Quite generally, the critical density for a given loop
size $s$ is
                            \begin{eq}{container-filled-with-hideous-rubbish}
\Dgreatercrit(s) = 1/s^{1/2} 
                                                                \end{eq}
per site. Here $\Dgreatercrit(s)$ counts loops equal to or bigger than
$s$, all of which contribute chain segments of length $s$ to the
density.  At the critical level their combined total mass ($s$ per loop
or strand) just fills a layer of thickness $s^{1/2}$, the equilibrium
size.  At higher densities the loops are stretched.

Returning to the many chain collapse issue, if all $1/N^{1/2}$ mother
loops per site were able to collapse unhindered, each would generate
$N/\sstar$ loops of size $\sstar$ by time $t$, giving a net density 
$\approx N^{1/2}/\sstar$.  But since this exceeds the critical level,
$1/{\sstar}^{1/2}$, it is clear the unhindered collapse costs energy and
must have been interfered with.

What then is the form of the loop distribution per site $\Omegat(s)$
at time $t$? To proceed, we make the following assumptions: (i) At
time $t$, loops of size $\sstar(t)$ are created on the surface up to
the critical density and no further, (ii) sizes greater than
$\sstar(t)$ follow a power law distribution and (iii) this
distribution is frozen in time. Assumptions (i), (iii) state that when
enough time has elapsed for a certain loop size to have been created,
creation of these loops proceeds but is then permanently switched off
as soon as other chains in the layer are forced to stretch to
accommodate further such loops.  Thus, equating $\int_{\sstar}^N ds
\,\Omegat(s)$ to $\Dgreatercrit(\sstar)$ determines $\Omega \approx
1/s^{3/2}$ for all $s>\sstar$ (self-consistently, the integral's lower
limit dominates).

Smaller loops, $s\ll\sstar(t)$, follow a different power law.  From
our single chain analysis we know these are unlikely to have been
created by time $t$ even without interference and must therefore be
very dilute.  For these, we can essentially repeat the single chain
arguments: eq. \eqref{loop-kinetics} (again dominated by the source
term) leads to $\Omegatdot \approx k(s) \Gammabound(t)$ where the
total density of loops $\Gammabound(t)\approx 1/{\sstar}^{1/2}(t)$ is
dominated by $\sstar$. The crucial point is that smaller loops are
sub-critical and follow free single chain kinetics, $k(s|\sstar)
\approx k(s)$. Their distribution thus follows the single chain power
law $\Omegat(s)\twid k(s) \twid 1/s^{1/2}$.  Demanding continuity at
$\sstar$ the overall layer distribution, shown in fig. \ref{scheme}(b),
is
                                                \begin{eq}{edible-car}
\Omegat(s) \approx
\casesshortii{c \, s^{-3/2}} {\sstar(t)<s<N}
        {  (c'/\sstar(t))  \, s^{-1/2}} {s<\sstar(t)} \period
                                                                \end{eq}
where $c,c'$ are constants of order unity. From eq. \eqref{edible-car}
the total density of loops bigger than $s$ is $\Dgreater(s) \approx
1/{\sstar}^{1/2}(t)$ for any small loop size $s<\sstar$.  This confirms that
such loops are sub-critical, an important self-consistency check.  A
sketch of the adsorbed layer loop structure is shown in
fig. \ref{blobs-melt}(a), characterised by a hierarchy of successively
smaller loops frozen in as they reach criticality.


\section{Guiselin experiment: swelling layer with solvent} 

Let us now ask what would happen in a Guiselin type experiment where
the chemisorption is interrupted after some time $\tfinal$ and the
brush subsequently swollen with good swelling solvent.  Typical values
of $Q$ are in the range \cite{ben:chemiphysi_letter} $10^{-3} \lsim Q
\lsim 10^2$ sec$^{-1}$ so surface saturation may not be reached for
minutes or hours. After interruption, the loop distribution is frozen
into the form given by eq. \eqref{edible-car}.  Neglecting
distinctions between tails and loops
\cite{semenovjoanny:loops_tails_europhys}, after solvent is introduced
the ``Guiselin brush'' structure is that of a polydisperse brush
having chain length distribution $\Omegat(s)$ evaluated at
$t=\tfinal$.

In ref. \cite{guiselin:irrev_ads} the density profile of such
polydisperse grafted layers was analysed, starting from the concept of
a local blob size \cite{gennes:book} $\xi(z)$ at height $z$ determined
by the local chain grafting density $\rho(z)=1/\xi^2(z)$. The
coarse-grained chain stretching is $dz/ds \approx \xi/g$ where there
are $g=\xi^{5/3}$ units per blob and the density profile is $c(z) =
\rho\, ds/dz$. The essential point is that only those chains (or
loops) bigger than $s$ can reach the height $z(s)$ so the effective
grafting density is $\rho(z) = \Dgreater(s)$. This immediately gives
$c(z) \approx {\Dgreater}^{2/3}(s)$, $z(s) = \int_0^s ds'
{\Dgreater}^{1/3}(s')$ and brush height $h=\int_0^N ds \Dgreater(s)$.

                                                \begin{figure}[t] 
\onefigure[width=9.9cm]{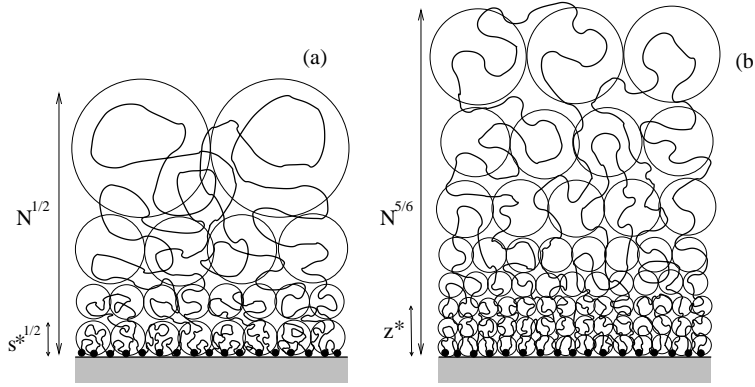}
\caption{ (a) Hierarchical loop structure of the adsorbed layer in the
melt. Successive loop size scales are frozen in as their critical
density is reached. Beyond this their formation would force chain
stretching and hence is strongly suppressed.  (b) Blob structure of the
layer swollen in good solvent. Blob
size is fixed $\approx {\zstar}^{3/10}$ in the inner region and
increases as $z^{3/10}$ in the outer layer.  }
                                                \label{blobs-melt}
                                                \end{figure}

For the present structure we have $\Dgreater(s) \approx s^{-1/2}$ for
large loops $s\gg\sstar$ and $\Dgreater(s) \approx {\sstar}^{-1/2}$ for
$s\ll \sstar$, valid provided $\sstar\ll N$. Thus
                                                \begin{eq}{fried-donut}
c(z) \approx \casesshortii{1/{\zstar}^{2/5} = (\Gammatot /h) \, (h/\zstar)^{2/5}}{z<\zstar}
               {1/z^{2/5} = (\Gammatot /h) \, (h/z)^{2/5}}{z>\zstar}
                        \comma\gap \zstar \equiv (Q\tfinal)^{-5/3} \comma
                                                                \end{eq}
where $\tfinal \le Q^{-1}$. Here $\Gammatot=\intinf ds \,s\Omegat(s)
\approx N^{1/2}$ is the total adsorbed mass per site and $\zstar
\equiv z(\sstar)={\sstar}^{5/6}$.  This profile, sketched in
fig. \ref{scheme}(c), has two distinct regions. (1) The inner
region of total mass $\Gammainner = \Gammatot (\sstar/N)^{1/2} \twid
1/\tfinal$, a small fraction of the total, has constant density
$\cinner =\Gammainner/\zstar \twid \tfinal^{2/3}$ up to $\zstar$.  (2)
The more diffuse outer part, from $\zstar$ to the brush height
$h=N^{5/6}$, where density decays as $z^{-2/5}$. This latter is the
same decay as for Guiselin's physisorbed brush.  Thus as the
chemisorption experiment is interrupted at later and later times
$\tfinal$ so the inner region becomes progressively thinner and
disappears at $\tfinal \approx 1/Q$.  The blob structure is shown
schematically in fig. \ref{blobs-melt}(b).


\section{Partially functionalised chains} 

When a fraction $f$ of chain units are chemically reactive ($f=1$ was
considered so far) there are now 2 monomer species whose relative
affinity for the surface is an essential new parameter. Suppose there
is an energy penalty $\epsilon$ when a reactive monomer displaces an
unreactive one adjacent to the surface.  Here we restrict attention to
weakly repulsive or attractive surfaces, $|\epsilon|<kT$.

In the simplest case of a {\em neutral} surface ($\epsilon=0$) most of
the previous discussion is unchanged provided one replaces $Q\gt fQ$
and the loop cascade is truncated at $\smin = f^{-1}$ at the
corresponding loop lifetime, \ie at $\tfinal = 1/(Qf^{1/2})$ (see
eq. \eqref{curried-rabbit}). That is, after coarse-graining over
scales $\smin$ one recovers the case of $100\%$ functionalization, but
with reduced effective reactivity $Qf$. The final loop density is
$\Gammaboundfinal = 1/\smin^{1/2}$ and the swollen Guiselin brush is
essentially unchanged, except that $\zstar$ reaches a minimum value
${\smin}^{5/6}$ for long chemisorption times.

This is actually a slight simplification; $\Gammaboundfinal$ cannot be
the true total surface loop density as $t\gt\infty$, since the
irreversibility of these reactions means that ultimately surface
coverage must reach unity, $\Gammabound\gt 1$. However, beyond
$\tfinal$ further coverage requires bringing down larger loops above
their critical density. We find the consequent stretching energy
penalty leads to exponentially suppressed reaction rates
\cite{ben:reactiface_pol_letter,ben:reactiface_pol} and a slow
logarithmic saturation $\Gammabound \approx \Gammaboundfinal
\ln^{1/2}(t/\tfinal)$.  Thus for very large reaction times we predict
the inner constant-density region ($z<\zstar$) of the swollen brush
will shrink and eventually disappear.

Consider now a surface weakly attractive to the reactive monomers,
$\epsilon<0$.  Their tendency to preferentially physisorb prior to the
much slower process of chemisorption is then weak; the entropic
disadvantage (giving free energy cost of order $kT$) of immobilization
at the surface is not worth the energy gain. Similar remarks apply to
weakly repulsive surfaces. In either case, the phenomenology is
unchanged from the neutral case after replacing $Q\gt
Qe^{-\epsilon/kT}$.

Let us also consider the more complex situations where only a fraction
$X$ of chains are functionalised, beginning with neutral surfaces.
The early stages of chemisorption now entail $\approx XQt$ single
chains adsorbing independently, each producing $L(t)$ loops of size
$\sstar(t)$ as described by eq. \eqref{loop-distribution} and
preceding text.  The net loop density is $\Gammabound \approx XQt
L(t)\twid t^3$ reaching criticality when loops of size $XN$ have
descended after time $\tsat=Q^{-1}(XN)^{-1/2}$. The surface layer of
thickness $(XN)^{1/2}$ is saturated and adsorption of new chains then
essentially halts.  The swelling experiment produces a Guiselin type
brush as described previously, but with $N$ replaced by $XN$.
Incomplete functionalization, $f<1$, is dealt with as before.

Non-neutral surfaces are more complex because segregation effects now
play a crucial role. A mean field estimate of the energy of a chain
close to the surface is proportional to the number of unperturbed
surface contacts $N^{1/2}$ times the probability a given contact is a
reactive monomer, $\Delta E \approx f N^{1/2} \epsilon$. When this is
below $kT$, \ie $f<\fcrit= N^{-1/2} (kT/\epsilon)$, the neutral
surface phenomenology is unmodified.  For more heavily functionalised
chains, chain configurations are strongly perturbed.  This case will
be discussed in ref. \cite{ben:chemiguiselin}.


\section{Topological Constraints}  
We end with a few remarks on how layer formation kinetics are
interfered with by topological constraints, frozen in as interwoven
loops are grafted to the surface. Thus far such effects were
neglected.  Consider for simplicity $f=1$ and neutral surfaces. At
time $t$, the layer is a network of grafted loops with characteristic
length $\sstar(t)$. This network, bathed in a ``solvent'' of ungrafted
free chains, has monomer density $\phinet \approx \Gammabound
{\sstar}^{1/2} \approx NX/\sstar$. Thus we expect topological
constraints to be unimportant provided $\sstar < \Ne$ where $\Ne =
\Nemelts/\phinet^{\gamma}$ is the entanglement threshold associated
with this density. Here $\gamma$ is an empirical system-dependent
exponent \cite{viovy:constraint_release_reorg_vs_dilation} and
$\Nemelts$ the value in the melt.  Insisting on this condition at all
stages during layer formation (from $\sstar = N$ at $t=0$ to $\sstar =
XN$ at saturation) we conclude that provided $N<\Nemelts/X$ (if
$\gamma>1$) or $N<\Nemelts/X^\gamma$ (if $\gamma<1$) then loops can
always explore all configurations and loop rate constants $k(s|s')$
are governed by Gaussian statistics as assumed in our picture.  If
this condition is satisfied, topological constraints are irrelevant
even after saturation, $t>\tsat$, since $\sstar$ then continues to
decrease while the monomer density of the non-frozen part of the
network ($s\le\sstar$) remains unity.  Hence this part remains
unentangled.

In summary, topological constraints do not interfere with the layer
kinetics provided either: (i) the melt is unentangled or (ii) if the
melt is entangled, the reactive chain fraction $X$ must be small
enough, $X<\Nemelts/N$ (taking a typical value $\gamma=1$). The local
reaction rate $Q$ must also be small enough to allow exploration of
all chain configurations before inhibition by further reactions.
Noting an unreacted loop of length $N$ makes of order $N^{1/2}$
surface contacts, the condition is $Q N^{1/2} T_N < 1$ where $T_N
\twid N^2$ is the loop Rouse relaxation time (for unentangled
melts\cite{gennes:book}) or $T_N \twid N^5$ (for entangled melts where
loops relax via constraint release
\cite{viovy:constraint_release_reorg_vs_dilation}).  If the conditions
on $N$ and $X$ are not satisfied, then no matter how small $Q$ at a
certain stage the entire layer freezes on scales large enough to be
entangled (with the exception of tails which can relax via arm
retraction mechanisms \cite{gennes:book}). These situations will be
discussed in a forthcoming publication \cite{ben:chemiguiselin}.


\section{Conclusions}  We have shown that slow formation of an
adsorbed surface layer from a melt leads to a modified version of the
Guiselin brush after swelling with solvent. In fact if the adsorption
process is allowed to proceed to completion ($\tfinal\gt\infty$) the
density profile of the swollen brush is unchanged from that predicted
by Guiselin for instantaneous adsorption.  For general interruption
times $\tfinal$, there appears a new inner region of constant density
whose width depends on $\tfinal$.

We have shown that reactions produce a characteristic loop
distribution at the surface consisting of two power laws for small and
large loops, respectively, with the dividing loop size $\sstar$
dependent on $\tfinal$.  Our analysis describes systems where some
or all chain units attach irreversibly to a surface but require many
collisions to ``cement in'' these attachments. This is important in
many applications involving chemisorption where the polymer loops
created by surface reactions serve as bridges enhancing interfacial
fracture toughness and yield stress after cooling
\cite{kraus:book_reinforcement_elastomers,wu:polymer_iface_adhesion_book,%
edwards:review_filler_reinforcement,cleggcollyer:book,%
pireaux:polymer_solid_ifaces_book}.  The resulting interfacial
strength depends strongly on loop size distributions
\cite{creton:kramer:iface_fracture_review,leger:anchored_chains_adhesion_review}.

\acknowledgments

This work was supported by the National Science Foundation, grant
no. DMR-9816374.


\begin{thebibliography}{10}

\bibitem{guiselin:irrev_ads}
Guiselin O., {\it Europhys. Lett.}, {\bf 17} (1992) 225--230.

\bibitem{auvray:irrev_ads_neutron}
Auvray L., Auroy P. and  Cruz M., {\it J. Phys. I France}, {\bf 2} (1992)
  943--954.

\bibitem{auvray:irrev_ads_concentrated_solution}
Auvray L., Cruz M. and  Auroy P., {\it J. Phys. II France}, {\bf 2} (1992)
  1133--1140.

\bibitem{ben:pmma_adsorption}
Durning C.~J., O'Shaughnessy B., Sawhney U., Nguyen D., Majewski J. and  Smith
  G.~S., {\it Macromolecules}, {\bf 32} (1999) 6772--6781.

\bibitem{marzolin:eng_graft_and_guiseling_brush}
Marzolin C., Auroy P., Deruelle M., Folkers J.~P., L\'{e}ger L. and  Menelle
  A., {\it Macromolecules}, {\bf 34} (2001) 8694--8700.

\bibitem{kraus:book_reinforcement_elastomers}
Kraus G., Reinforcement of Elastomers (John Wiley \& Sons, New York, 1965).

\bibitem{wu:polymer_iface_adhesion_book}
Wu S., Polymer Interface Adhesion (Marcel Dekker, New York, 1982).

\bibitem{edwards:review_filler_reinforcement}
Edwards D.~C., {\it J. Mater. Sci.}, {\bf 25} (1990) 4175--4185.

\bibitem{cleggcollyer:book}
Mechanical Properties of Reinforced Thermoplastics (Elsevier, London, 1986).
\newblock Edited by D. W. Clegg and A. A. Collyer.

\bibitem{pireaux:polymer_solid_ifaces_book}
Polymer-Solid Interfaces (Institute of Physics Publishing, Bristol, 1992).
\newblock Edited by J. J. Pireaux, P. Bertrand and J. L. Bredas.

\bibitem{creton:kramer:iface_fracture_review}
Creton C., Kramer E.~J., Brown H.~R. and  Hui C.-Y., {\it Adv. Pol. Sci.}, {\bf
  156} (2001) 53--136.

\bibitem{thon:rateconstants_book}
Tables of Chemical Kinetics. Homogeneous Reactions (National Bureau of
  Standards, Department of Commerce, 1951).
\newblock Edited by Thon, N.

\bibitem{ben:chemiphysi_euro}
O'Shaughnessy B. and  Vavylonis D. (Eur. Phys. J. E, in press,
  cond-mat/0301206).

\bibitem{ben:chemiguiselin}
O'Shaughnessy B. and  Vavylonis D. (in preparation).

\bibitem{shafferchakraborty:pmma_chemisorption_kinetics}
Shaffer J.~S. and  Chakraborty A.~K., {\it Macromolecules}, {\bf 26} (1993)
  1120--1136.

\bibitem{ben:chemiphysi_letter}
O'Shaughnessy B. and  Vavylonis D., {\it Phys. Rev. Lett.}, {\bf 90} (2003)
  056103.

\bibitem{semenovjoanny:loops_tails_europhys}
Semenov A.~N. and  Joanny J.-F., {\it Europhys. Lett.}, {\bf 29} (1995)
  279--284.

\bibitem{gennes:book}
{de Gennes} P.~G., Scaling Concepts in Polymer Physics (Cornell Univ. Press,
  Ithaca, New York, 1985).

\bibitem{ben:reactiface_pol_letter}
O'Shaughnessy B. and  Sawhney U., {\it Phys. Rev. Lett.}, {\bf 76} (1996)
  3444--3447.

\bibitem{ben:reactiface_pol}
O'Shaughnessy B. and  Sawhney U., {\it Macromolecules}, {\bf 29} (1996)
  7230--7239.

\bibitem{viovy:constraint_release_reorg_vs_dilation}
Viovy J.~L., Rubinstein M. and  Colby R.~H., {\it Macromolecules}, {\bf 24}
  (1991) 3587--3596.

\bibitem{leger:anchored_chains_adhesion_review}
L\'{e}ger L., Rapha\"{e}l E. and  Hervet H., {\it Adv. Pol. Sci.}, {\bf 138}
  (1999) 185--225.

\end{thebibliography}

\end{document}